\documentclass[fleqn,usenatbib]{mnras}

\usepackage{siunitx}
\usepackage[version=4]{mhchem}
\usepackage{multirow}
\usepackage{graphicx}
\usepackage[caption=false]{subfig}
\usepackage[strict]{changepage}
\usepackage{xcolor}
\usepackage{ulem}
\usepackage{amssymb}
\usepackage{amsmath}
\usepackage{lscape}
\usepackage{mathtools}
\usepackage{natbib}
\usepackage{pdflscape}
\usepackage{gensymb}

\usepackage{tikz}
\usetikzlibrary{shapes.geometric, arrows.meta, positioning, fit, backgrounds}
\defcitealias{zhou2023}{Paper~I}   

\DeclareSIUnit\jansky{Jy}

\newcolumntype{L}[1]{>{\raggedright\let\newline\\\arraybackslash\hspace{0pt}}m{#1}}
\newcolumntype{C}[1]{>{\centering\let\newline\\\arraybackslash\hspace{0pt}}m{#1}}
\newcolumntype{R}[1]{>{\raggedleft\let\newline\\\arraybackslash\hspace{0pt}}m{#1}}

\begin{document}


\title[Automated void identification by Blendmask]{Automated void identification by Blendmask: from hierarchical molecular gas to hierarchical voids in NGC 628}

\author[J. W. Zhou]{
J. W. Zhou \thanks{E-mail: jwzhou@mpifr-bonn.mpg.de}$^{1}$
A. A. Han $^{2}$
\\
$^{1}$Max-Planck-Institut f\"{u}r Radioastronomie, Auf dem H\"{u}gel 69, 53121 Bonn, Germany\\
$^{2}$State Key Laboratory of Resources and Environmental Information System, Institute of Geographic Sciences
and Natural Resources Research,\\ Chinese Academy of Sciences, Beijing 100101, China
}

\date{Accepted XXX. Received YYY; in original form ZZZ}
\pubyear{2025}
\maketitle

\begin{abstract}
We identify voids in NGC 628 from the JWST MIRI F770W image using a deep-learning method (BlendMask) and refine them by intensity contrast. These voids may be feedback-driven bubbles or dynamically formed structures. Cross-matching with archival catalogs of star clusters and associations shows that only up to 17.6\% of voids are associated with such stellar populations. The HST B-band peak-flux distributions of voids with and without these populations overlap substantially, suggesting that many related clusters/associations remain unidentified or misclassified in current catalogs.
Voids associated with star clusters or associations tend to have lower intensity contrast and larger sizes. An anti-correlation between void size and intensity contrast indicates that larger voids have emptier centers, possibly due to more substantial feedback. Hence, voids may provide a complementary tracer for identifying stellar populations and constraining their physical properties.
To quantify spatial relationships among CO, 21$\mu$m, H$_{\alpha}$ sources, and voids, we construct networks linking each source pair. Among the nine networks, 21$\mu$m and H$_{\alpha}$ sources show the strongest spatial association. Compared to small voids, 
large voids exhibit progressively increasing separations from CO to 21$\mu$m, then to H$_{\alpha}$ sources, and finally to the voids,
consistent with an evolutionary sequence in space and time. Smaller voids lie closer to molecular clouds, while larger voids are more displaced. 
Compared with molecular clouds not associated with voids, those associated with voids are significantly more massive and appear to be more evolved. In fact, 68\% of molecular clouds associated with voids are also associated with 21$\mu$m sources. 
These results support an evolutionary scenario in which some voids originate within molecular clouds, grow through stellar feedback, and gradually detach from their parent clouds.
\end{abstract}


\maketitle

\section{Introduction}\label{intro}

Galaxies act as stellar factories in the universe, producing stars through the gravitational collapse of the densest regions within molecular clouds in the interstellar medium (ISM). Molecular clouds are widely distributed within galaxies and serve as localized sites for star formation. Each molecular cloud typically contains multiple clumps, which are the local sites of star formation within the cloud and the precursors of embedded star clusters \citep{Miville2017-834,Rosolowsky2021-502,Urquhart2022-510,Yan2017-607,Zhou2024PASP-1,Zhou2024PASP-2}.
As presented in \citet{Motte2018-56, Vazquez2019-490,Kumar2020-642, Henshaw2020-4, Zhou2022-514,Zhou2023-676,Zhou2024-686-146, Zhou2024PASA, Zhou2024-534,Zhou2025-537-2630}, the molecular gas from galaxy to dense core scales presents multi-scale/hierarchical hub-filament structures. 
As a dense core serves as the hub within a clump, and a clump as the hub within a molecular cloud, a molecular cloud can be the hub within a galaxy.
Hubs at different scales as local gravitational centers can continuously accrete surrounding diffuse gas, thereby accumulating mass.
The hierarchical structure of molecular gas in a galaxy determines the spatial distribution of stellar populations \citep{Gouliermis2015MNRAS-452, Elmegreen2014-787,Grasha2017-840,Chevance2023-534}. 
Young clusters with different ages constitute a cluster population within the parent molecular cloud. The evolution of the molecular cloud is closely related to the formation and evolution of the internal star clusters. The star clusters become fully exposed after the molecular cloud is dispersed by feedback from the star clusters \citep{Kim2016-819,Kruijssen2019-569,Chevance2022-509,Watkins2023-676}.

The evolution of galaxies and the state of the ISM are heavily influenced by the activity of massive stars. Through powerful stellar winds, intense ionizing radiation, dust heating, and the expansion of HII regions, they inject energy into their surroundings and cause significant dynamical disturbances. These processes create bubbles in the ISM. Over time, individual bubbles may merge, and the additional input from supernova (SN) explosions can amplify the effect, leading to the formation of larger structures called superbubbles \citep{Mac1988-324,Churchwell2006-649,Zinnecker2007-45, Dale2012-424, Krumholz2014-243K, Rahner2017-470,Kim2018-859,Grasha2019-483,Nath2020-493,Barnes2021-508,Jeffreson2021-505,Lewis2023-944,Zhao2024-974}. 
Although stellar feedback plays a crucial role in regulating galaxy evolution, observational constraints on the relative contributions of different feedback processes—and their role in disrupting molecular clouds—remain incomplete \citep{Matzner2002-566,Walch2012-427, Rogers2013-431, Dale2014-442, Lopez2014-795, Matzner2015-815, Smith2018-478, Rugel2019-622, Watkins2019-628A, Barnes2022-662,Zakardjian2023-678,Zhou2024-682-128}. HII regions and bubbles act as vital indicators of feedback activity, providing valuable clues about supernova events, gas dispersal timescales, and whether feedback leads to the removal or destruction of gas, thus influencing star formation either positively or negatively. Gaining a deeper understanding of these mechanisms is crucial for revealing how bubbles form and evolve. This, in turn, helps us reconstruct the star formation histories of galaxies, trace the lifecycle of the ISM, and better understand the physical processes that regulate star formation rates \citep{Hopkins2014-445,Girichidis2016-456,Rathjen2021-504,Watkins2023-676,Schinnerer2024-62,Leroy2025-985}.

Void- and shell-finding in the ISM has traditionally relied on H I observations, leading to several large catalogs of holes and supershells in nearby galaxies and the Milky Way \citep{Heiles1979-229,Puche1992-103,Kim1998-503,Ehlerova2005-437,Bagetakos2011-141,Ehlerova2013-550,Pokhrel2020-160}.
More recent efforts have expanded both the datasets and the methodologies, which marks a transition from atomic-gas void studies toward multi-phase ISM structure identification across H I, CO, and dust tracers.
A prominent feature in mid-infrared (MIR) JWST observations of nearby galaxies is the presence of numerous voids in the ISM, ranging in size from parsecs to kiloparsecs \citep{Barnes2023-944,Watkins2023-944}.
The galaxies are densely filled with ring-like and shell-like structures, which may be created by feedback processes or may be dynamically driven features.
\citet{Watkins2023-944} manually identified bubbles in NGC 628 by combining Physics at High Angular resolution in Nearby GalaxieS
(PHANGS)$-$JWST MIRI F770W imaging with PHANGS$-$HST B$-$band and PHANGS$-$MUSE H$_{\alpha}$ data. As a result, it is undoubtedly that some bubbles were missed. Moreover, based on extrapolation of the NGC 628 results to the remaining 18 PHANGS$-$JWST galaxies, they estimated that an additional 37000 to 79000 bubbles could exist across the survey. This underscores the necessity of employing machine learning techniques to manage and analyze large-scale observational datasets effectively. In particular, deep learning offers both speed and accuracy in detecting complex structures, capturing subtle morphological features, and uncovering physical insights into bubble formation, evolution, and their interaction with the surrounding ISM \textcolor{magenta}{\citep{Van2019-880,Xu2020-890,Nishimoto2025-77}.}

In this work, based on the PHANGS$-$JWST MIRI F770W image, we develop robust methods for identifying voids in NGC 628. We then further investigate the physical mechanisms responsible for their formation and evolution, as well as assess their influence on star formation. 
NGC 628 with a distance of $\sim$9.84 Mpc is a face-on spiral galaxy. The inclination angle of NGC 628 is $\sim$8.9 degree \citep{Lang2020-897}. This facilitates the identification of voids embedded within the galaxy. Previous work \citep{Zhou2024-534,Zhou2025-537-2630} have systematically studied CO, 21$\mu$m, and H$_{\alpha}$ sources (intensity peaks) in NGC 628, which may have an evolutionary relationship with the voids identified in this work and are helpful for the following discussion. Additionally, \citet{Barnes2023-944,Watkins2023-944} have studied the bubbles in NGC 628, providing a reference for this research. 


\section{Data} 

For NGC 628, we used the moment-0 map derived from the combined 12m+7m+TP PHANGS-ALMA CO (2$-$1) data cube to investigate the molecular gas distribution, achieving a linear resolution of $\sim$50 pc at a distance of 9.8 Mpc \citep{Leroy2021-257, Jang2014-792, Kreckel2017-834}.
We also utilized maps from the PHANGS–JWST survey at 3.3, 7.7, and 21$\mu$m, with corresponding linear resolutions of $\sim$6, $\sim$11, and $\sim$30 pc, respectively. In addition, we incorporated the H$_{\alpha}$ emission map from the PHANGS–MUSE survey ($\sim$40 pc resolution) and the B-band emission map from the PHANGS–HST survey.
Overviews of the PHANGS-ALMA, PHANGS-MUSE, PHANGS-JWST and PHANGS-HST surveys' science goals, sample selection, observation strategy, and data products are described in \citet{Leroy2021-257,Leroy2021-255,Lee2022-258,Emsellem2022-659,Lee2023-944,Williams2024-273}.
In particular, \citet{Williams2024-273} provides a detailed description of the JWST data processing.
All the data are available on the PHANGS team website \footnote{\url{https://sites.google.com/view/phangs/home}}.

To facilitate a direct comparison with the bubble identification in \citet{Watkins2023-944} and to make use of their catalog as a training set, our analysis is primarily based on the MIRI F770W image. As discussed in \citet{Watkins2023-944}, the F770W image provides high-resolution polycyclic aromatic hydrocarbons (PAHs) emission that strongly enhances the contrast of bubble (void) shells while minimizing contamination from stellar continuum emission,
allowing small-scale structure identification.
Note that we identify the voids purely based on the F770W image, without distinguishing between feedback-driven bubbles and dynamically driven features during the identification process.

\section{Results and discussion} 

\subsection{Void identification} 

\subsubsection{The deep learning method}\label{blendmask}

\begin{figure}
\centering
\includegraphics[width=0.475\textwidth]{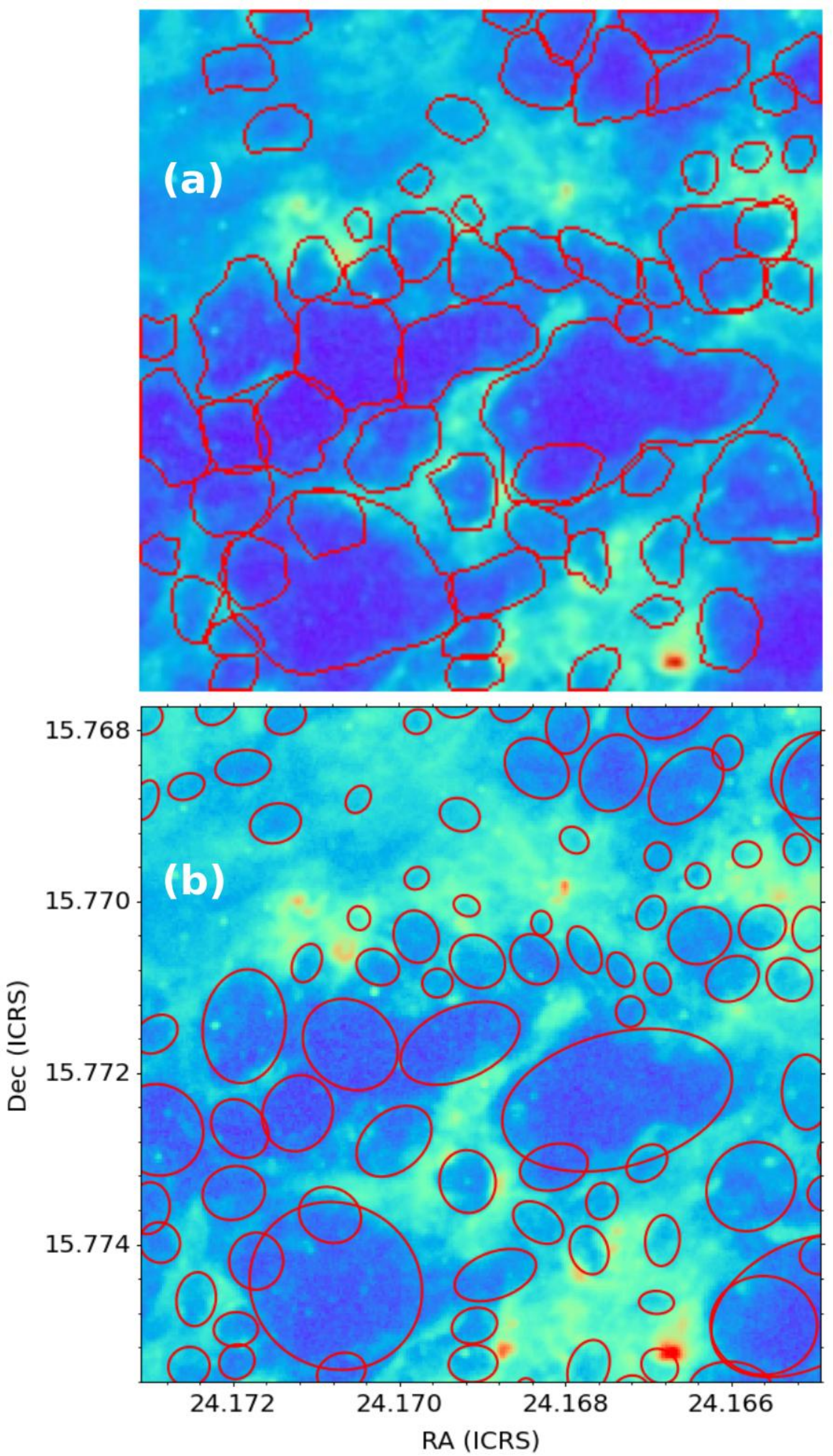}
\caption{The JWST F770W image of a subregion in NGC 628 marked by the blue box in Fig.\ref{ML-small}.
(a) Red contours marked by LabelMe show the voids recognized by the human eye; (b) BlendMask outputs segmentation masks for all detected objects. We fit ellipses to the mask boundaries to extract center coordinates, major/minor axis lengths, and tilt angles. 
}
\label{label}
\end{figure}

\begin{figure}
\centering
\includegraphics[width=0.475\textwidth]{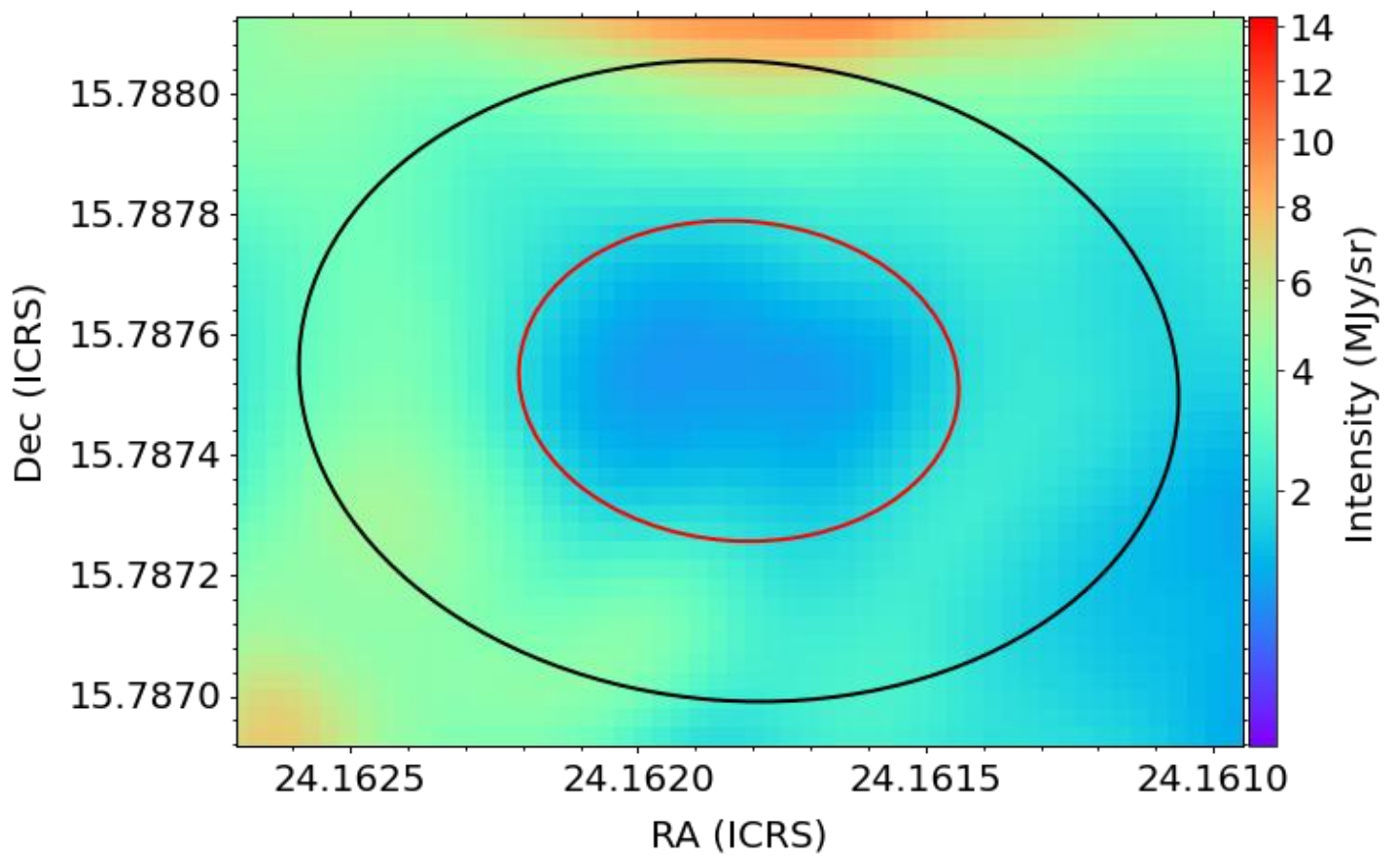}
\caption{A void used to show the calculation of the intensity contrast in Sec.\ref{c}. 
}
\label{contrast}
\end{figure}

\begin{figure*}
\centering
\includegraphics[width=1\textwidth]{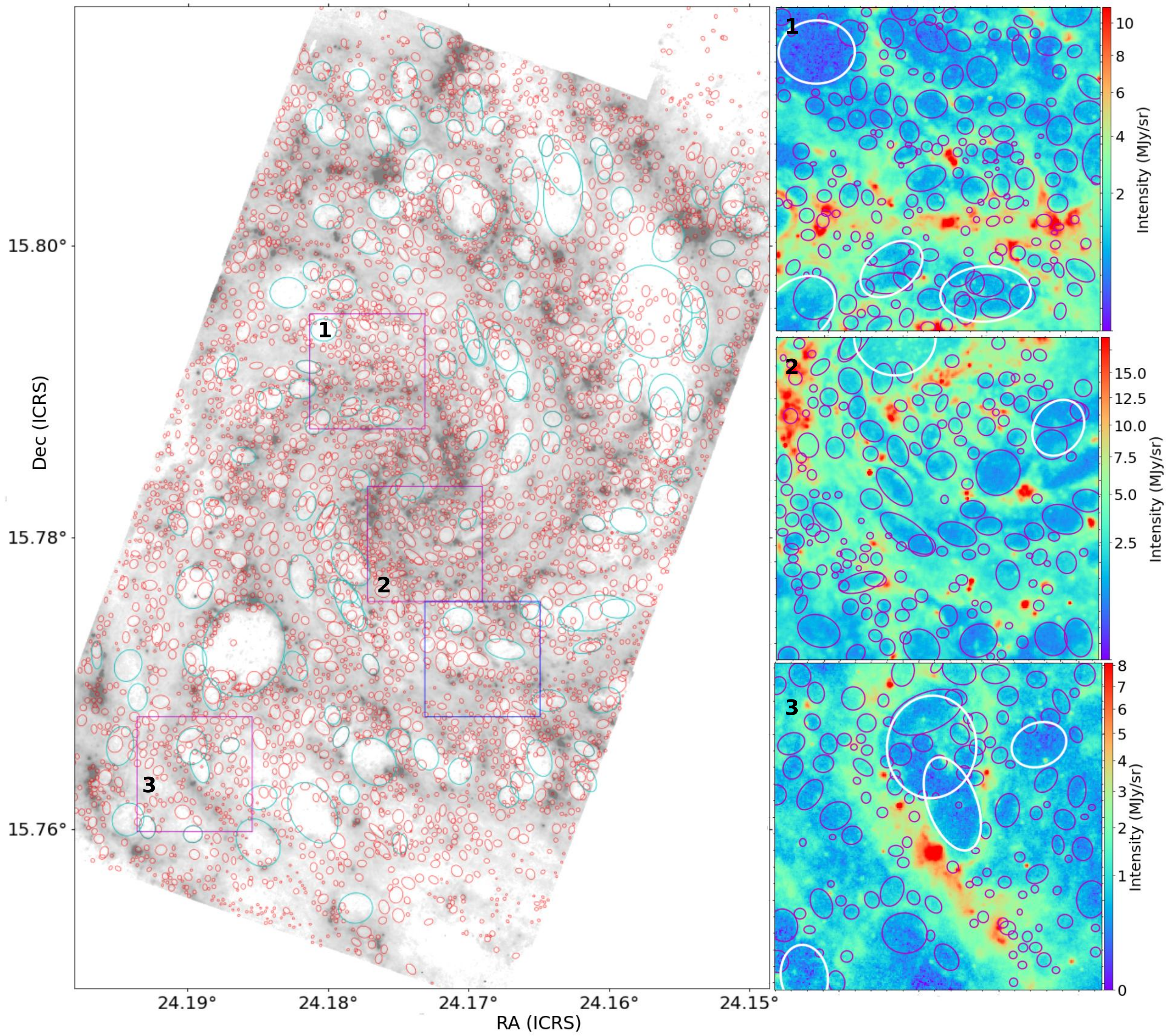}
\caption{Red ellipses denote small-scale voids ($<120$~pc), while cyan dashed ellipses indicate large-scale voids ($>120$~pc) in the merged catalog, after removing hierarchical voids as described in Sec.\ref{layer}. The three magenta boxes mark the zoom-in regions shown on the right (magenta ellipses: $<120$~pc; white ellipses: $>120$~pc), and the blue box outlines the area displayed in Fig.\ref{label}. The background of all maps is the JWST F770W image.}
\label{ML-small}
\end{figure*}

In the JWST F770W image, the entire galaxy is filled with voids. To improve the efficiency of structure identification and explore a method suitable for large-sample structure recognition, we explore the use of deep learning for void identification.
Identifying voids requires separating them from the background, which can be transformed into an instance segmentation task. 
The Mask Region-based Convolutional Neural Network (Mask R-CNN) framework \citep{He2017Mask} for instance segmentation has been widely used in astronomy for source detector and structural identification \citep{Burke2019-490,Ahmadzadeh2019,Farias2020,Alina2022,Lao2023,Riggi2023,Liang2023}. It creates bounding boxes and segmentation masks for each object instance in the image. In other words, it identifies each object in an image and provides details about its location. 
BlendMask \citep{Chen2020BlendMask}, as an instance segmentation model, enhances the Mask R-CNN framework with several modifications to improve accuracy and efficiency. It combines the strengths of Mask R-CNN with an attention mechanism and a feature blending module to enhance the segmentation process. Specifically, instead of predicting each instance mask independently within region proposals as in Mask R-CNN, BlendMask first constructs a shared dense feature map and then generates instance-specific attention maps from detection features. These attention maps are used to selectively weight and blend the shared features, producing the final instance masks in a more efficient manner. This design reduces redundant per-instance computation while preserving fine spatial details, and is particularly beneficial for structures with irregular shapes and diffuse boundaries, such as voids.
Therefore, we use BlendMask to perform the instance segmentation task in this work.

To better visualize small-scale voids while simultaneously preserving large-scale voids, we used a sliding window to divide the original large image into multiple smaller patches. 
After testing a range of window sizes, we adopted a final window size of
256 pixel $\times$ 256 pixel ($\sim$ 1343 pc $\times$ 1343 pc), with a step size of 128 pixels. Voids can be easily recognized by the human eye in the image, then we use LabelMe \citep{LabelMe} to annotate all voids within selected patches as the training set.
LabelMe is a popular image annotation tool, especially for computer vision tasks, such as object detection and segmentation. LabelMe creates a JSON file for each image, which contains detailed annotation information (such as coordinates, categories, etc.). Then, the JSON file from LabelMe needs to be converted into a format suitable for machine learning frameworks, for example, converting the JSON file to the COCO format for object detection. After that, a deep learning framework like PyTorch can be used to perform object detection or image segmentation tasks \citep{Quan2021,Zhao2022RemS,Su2023Senso,Xu2024RemS}. 
Finally, in $\sim$170 patches, 10 independent, non-overlapping patches located in different regions of the galaxy were used as the training set.
One example displays in Fig.\ref{label}(a).

In terms of BlendMask's capabilities, void identification is a very simple task. 
Due to the simple structure and morphological similarity of voids, it is not necessary to annotate every instance. Labeling representative examples is sufficient, as the model can automatically identify unannotated voids of the same type. The deep learning model has a certain degree of fault tolerance, so a portion of inaccurate labels does not significantly affect the performance of machine learning, especially when the sample size is large. There are differences between the structures marked in Fig.\ref{label}(a) and (b), because the final identification results depend on the model's performance learned from the entire training set, not just a single patch.
In summary, our task is simple, and the structures to be recognized are also simple. Therefore, the operation of BlendMask is robust.
Moreover, after the structure identification, we further filter the identified structures by the intensity contrast defined in Sec.\ref{c}.

The output of BlendMask typically consists of segmentation masks for each detected object, along with associated confidence scores.
We perform ellipse fitting on the boundary of each mask to obtain the center coordinates, the lengths of the major and minor axes, and the tilt angle of the equivalent ellipses, as shown in Fig.\ref{label}(b). 
Since there is significant overlap between adjacent patches, the identified structures also exhibit substantial repetition. We adopted two criteria to eliminate duplicate structures: (1) The distance between the centers of two equivalent ellipses is less than than one PSF (point spread function) FWHM ($\sim$0.67$''$); (2) The ratio of the absolute difference in area between two equivalent ellipses and the total area of the smaller ellipse is less than 1. Finally, 2809 structures are retained (catalog-1).

For comparison with manual annotation by LabelMe, we directly used the bubble catalog of \citet{Watkins2023-944}  (catalog-2) as the training set. 
In \citet{Watkins2023-944}, the bubbles were identified by eye from multi-wavelength imaging, but these constitute only a subset of all voids within the galaxy. We selected the 10 patches with the highest bubble density to reduce the impact of incomplete sample construction, since bubbles represent only a subset of the voids. 
For the newly selected patches, only one overlaps with the patches previously used for the training set. 
Following the above procedure, finally, 2659 structures are retained (catalog-3).
From a morphological point of view, voids are all very similar, regardless of whether they are bubbles produced by feedback. Even if the training set includes only the bubbles in the catalog of \citet{Watkins2023-944}, the algorithm will automatically identify many other voids that are not in the catalog.

\subsubsection{Intensity contrast}\label{c}

The typical feature of a void is that it is almost empty in the middle. This is exactly opposite to the hub-filament structure presented in \citet{Zhou2024-534}. As illustrated in Fig.\ref{contrast}, by analogy with the density contrast defined in \citet{Zhou2024-534}, we can also define an intensity contrast for voids, i.e. the ratio of the average intensity in the central region of a void marked by the red ellipse in Fig.\ref{contrast}(a), $F_{\rm in}$, to the average intensity within an elliptical ring around the central region of the void (the region between the black and red ellipses in Fig.\ref{contrast}(a)), $F_{\rm out}$,
\begin{equation}
C=F_{\rm in}/F_{\rm out}. 
\end{equation}  
The width of the ring is equal to the equivalent radius of the central identified void (the square root of the product of the semi-major and semi-minor axes of the equivalent ellipse).
A real void should satisfy $C<1$. 
For the above catalog, the numbers of voids that satisfy this condition are 2702 (catalog-1), 1585 (catalog-2), and 2597 (catalog-3), respectively.

Among them, catalog-1 and catalog-2 share 874 bubbles; catalog-3 and catalog-2 share 967 bubbles; and catalog-1 and catalog-3 together share 1190 bubbles (75\%) with catalog-2.
With regard to catalog-3, although we used only 10 patches, we recovered 61\% of the bubbles in catalog-2, demonstrating the effectiveness of the algorithm.
Then we combined the three catalogs and applied the same procedure to remove duplicate structures, resulting in a final set of 5441 voids, as shown in Fig.\ref{ML-merge}. To verify the completeness of the voids in the merged catalog, we randomly selected 10 patches again and used the voids from the merged catalog as the training set. After removing duplicate structures, we obtained 4877 voids, of which 4873 are present in the merged catalog. We therefore conclude that the merged catalog is complete.

\subsection{Hierarchical voids}\label{layer}

In Fig.\ref{ML-merge}, there are many locally overlapping voids. Since we have removed repeated voids, these overlapping voids actually reflect the hierarchical structure among voids at different scales, which is also presented in \citet{Watkins2023-944}.
According to the physical picture described in Sec.\ref{intro}, the hierarchical structure of molecular gas in the galaxy determines the spatial distribution of stellar populations. Hierarchical star formation or young stellar populations \citep{Gouliermis2015MNRAS-452, Elmegreen2014-787,Grasha2017-840,Adamo2017-841} further indicate the presence of hierarchical feedback-driven bubbles.
However, in the following analysis, we need to eliminate these locally nested structures and retain only the lowest-level, mutually independent ones. 

To achieve this goal, the overlap fraction between each ellipse pair is computed. Pairs are only considered if their centers are closer than the maximum ellipse radius. 
Large-scale voids typically contain many small-scale voids. To avoid excessively removing small-scale voids, we divide the initial voids into two categories — large and small (120 pc as the boundary) — and process them separately.
Voids with a scale larger than 120 pc account for only 4\% of the total and are sparsely distributed within the 120–560 pc range.
In the small void category, if the overlap exceeds 75\%, we exclude large voids that overlap with small voids. Conversely, in the large void category, we exclude small voids that overlap with the large ones.
Fig.\ref{ML-small} shows the remaining 4594 voids with $C<1$. 
A comparison with the bubble catalog of \citet{Watkins2023-944} is presented in Fig.\ref{ML-cp}.

As can be seen in Fig.\ref{ML-small}, large-scale and small-scale voids still overlap, since they were processed separately. Voids may evolve from small to large scales over time. This work primarily examines the correlation between voids and their potential precursors—namely CO, 21~$\mu$m, and H$_\alpha$ sources identified in \citet{Zhou2024-534}. 
These precursors, which appear as intensity peaks, are comparable in size to the small-scale voids. Therefore, we focus only on the small-scale voids in the following analysis.
There are 4498 out of 4594 (98\%) voids with a radius less than 120 pc.

In Fig.\ref{ML-small}, 
some large-scale bubbles do not match well with the background image.
A possible reason is that the training set mainly consists of cropped patches, and some large-scale voids are comparable in size to an entire patch. In addition, the number of large-scale voids is much smaller than that of small-scale voids. These factors together make it difficult for the model to effectively learn the features of large-scale voids. Alternatively, the identification of large-scale voids may simply fall outside the scale range of interest in this work.
Since our analysis is based on patch-based processing, both in training and inference, large-scale voids are effectively suppressed. A large void, when occupying a significant fraction of a patch, may not be recognized as a distinct structure due to the lack of sufficient background contrast. Moreover, large-scale voids are relatively rare and can often be easily identified by visual inspection; thus, machine learning is not essential for their detection.
The primary strength of the machine learning approach lies in identifying the numerous small-scale voids, which are densely distributed and well suited for automated recognition. Therefore, we do not further discuss optimization of large-scale void identification, and they are also excluded from the subsequent analysis.

\subsection{Stellar populations within voids}\label{stellar}

\begin{table*}
\centering
\caption{Classification of voids based on their association with stellar populations. void-A0 is a subset of void-A. void-B1 and void-B2 are two subsets of void-B (void-B = void-B1 + void-B2).}
\label{tab1}
\begin{tabular}{cc}
\hline
Categories & Definition \\
\hline
void-A  & Voids associated with star clusters or associations, with a center separation smaller than the void equivalent radius. \\
void-A0 & Subset of void-A with a center separation smaller than half of the void equivalent radius. \\
void-B  & Voids not associated with star clusters or associations under the void-A criterion. \\
void-B1 & Subset of void-B with HST B-band peak fluxes above the minimum value of void-A. \\
void-B2 & Subset of void-B with HST B-band peak fluxes below the minimum value of void-A. \\
\hline
\end{tabular}
\end{table*}

\begin{figure*}
\centering
\includegraphics[width=0.925\textwidth]{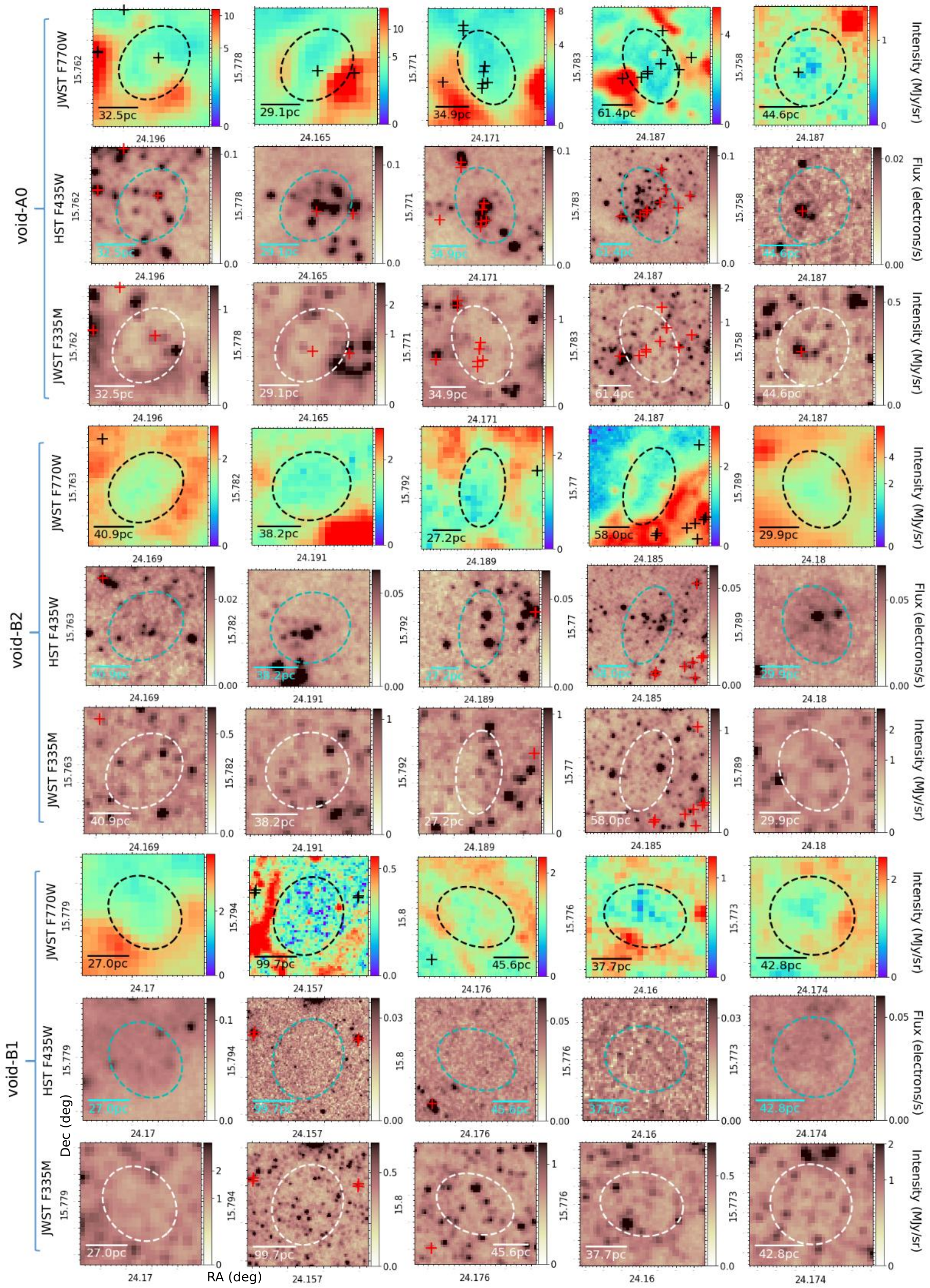}
\caption{Examples of voids from categories voids-A0, voids-B2, and voids-B1 classified in Sec.\ref{stellar}.
Table.\ref{tab1} summarizes the definitions of the different void categories.
For the five voids selected as examples from each category, the first, second and third rows show the JWST F770W, HST F435W and JWST F335M images, respectively. Ellipses indicate the identified voids, and pluses mark the central positions of the identified clusters or associations in the PHANGS-HST and the LEGUS surveys.}
\label{case}
\end{figure*}

Voids are identified solely based on the F770W images. These voids may correspond to either feedback-driven bubbles or dynamically driven structures.
We then further classify them by matching with stellar populations. Table.\ref{tab1} summarizes the definitions of the different void categories.

\subsubsection{Archived star cluster and association catalogs}

\citet{Maschmann2024-273} presents photometric catalogs for the largest census to-date of $\sim$100000 star clusters and compact associations resulting from the PHANGS-HST Treasury survey of 38 nearby spiral galaxies \citep{Lee2022-258}, including NGC 628.
Based on the classification technique, there are two kinds of catalogs, i.e. human classified and machine learning (ML) classified.
The ML classified technique greatly increases the number of identified star clusters and associations within each galaxy, but some members may be unreliable \citep{Maschmann2024-273}.
As a conservative estimate, we consider all identified star clusters and associations, whether human or ML classified. 
Additionally, we also incorporate the star cluster and association catalog of NGC 628 from the LEGUS (Legacy ExtraGalactic UV Survey) survey \citep{Adamo2017-841}. 
In these catalogs, sources classified as class 1, 2, or 3 are considered star clusters or associations.
The number of star clusters and associations in the human- or ML-classified catalogs of \citet{Maschmann2024-273} and the catalog of \citet{Adamo2017-841} is 789, 4750, and 1276, respectively. 
Although there is some overlap among the three catalogs, since our goal is merely to assess whether voids are associated with star clusters or associations, duplicate removal is not required.
While the catalog of \citet{Maschmann2024-273} does not provide physical properties of the clusters (e.g., age and mass), the LEGUS catalog \citep{Adamo2017-841} does, but with large uncertainties.

Of the 4498 voids, 4080 fall within the field of view of the HST observations.
Then we match star clusters and associations with voids by requiring that the separation between their central coordinates is smaller than the void's equivalent radius.
Finally, only 719 (17.6\%) voids are associated with star clusters or associations ("void-A"). If void formation is driven by stellar feedback, star clusters or associations are expected to be located near the centers of the voids. And those found near the edges of the voids may result from projection effects or triggered star formation. Moreover, as a conservative estimate, we also did not impose any constraints on the ages of star clusters and associations. Therefore, the 17.6\% value should be regarded as an upper limit. 
If we require that the distance between central coordinates be less than half of the void’s equivalent radius, the number drops from 719 to 261 ("void-A0"). 
HST B-band (blue) observations were used in \citet{Watkins2023-944} to identify feedback-driven bubbles. 
Only 719 voids are associated with star clusters or associations, which is even significantly fewer than the 1694 bubbles listed in the catalog of \citet{Watkins2023-944}.
These results indicate that most of the voids may not originate from stellar feedback, or that a large number of star clusters and associations in the galaxy are not in the current catalogs. Another possibility is that stellar associations are thought to become unbound on timescales comparable to those of bubble growth \citep{Gouliermis2015MNRAS-452, Elmegreen2014-787,Grasha2017-840,Adamo2017-841}, so it is not necessarily expected that all original stellar feedback sources would still be present.

\begin{figure}
\centering
\includegraphics[width=0.375\textwidth]{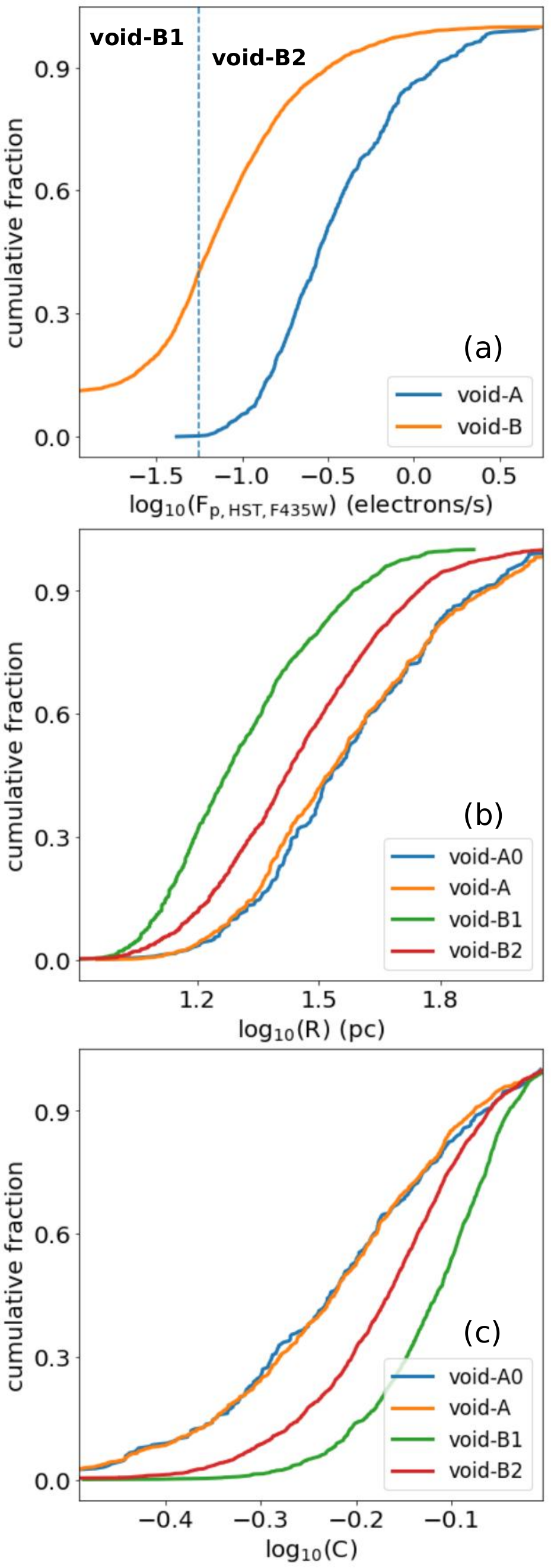}
\caption{Comparison of the physical parameters among different void categories.
(a) The HST B-band peak flux;
(b) Intensity contrast ($C$);
(c) Radius ($R$).
In panel (a), the dashed line divides void-B into two subsets. 
}
\label{compare}
\end{figure}

Fig.\ref{compare}(a) shows the HST B-band peak flux distribution of voids associated ("void-A") or not associated ("void-B") with star clusters or associations. We can see a significant overlap between the two categories, which supports that a large number of star clusters and associations associated with voids are not identified in current star cluster and association catalogs. Some cases are presented in Fig.\ref{case}.
As marked in Fig.\ref{compare}(a), using the lowest HST B-band peak flux of the voids in void-A as the standard, we divide void-B into two subsets, void-B1 and void-B2, containing 1550 and 2325 voids, respectively. A significant fraction of the voids in void-B2 are expected to be associated with star clusters or associations. For the star cluster and association catalog of NGC 628, further investigation is required to improve its completeness.
The analysis presented here suggests that voids can aid in the identification of star clusters and associations and may also serve as a tool to constrain their physical parameters.
Morphologically, the voids in void-B1, void-B2 and void-A show no significant differences, as illustrated in Fig.\ref{case}. Given the current observational evidence, we cannot conclude that the voids in void-B1 are not from feedback-driven. Certainly, some of them may simply be dynamically driven features within the galaxy.


\begin{figure}
\centering
\includegraphics[width=0.475\textwidth]{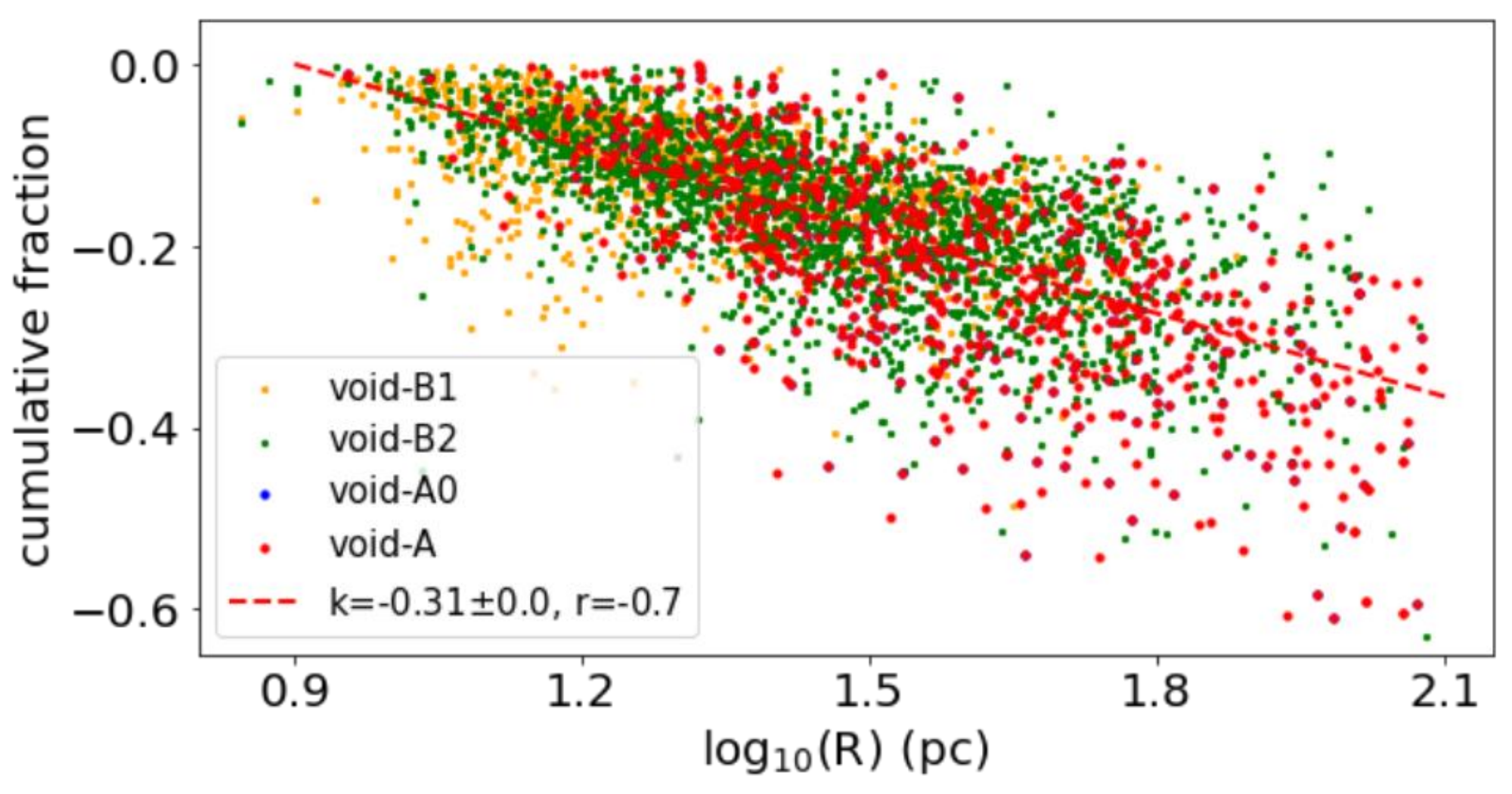}
\caption{Correlation between the intensity contrast and radius of voids.
$k$ and $r$ are the slope and the Pearson correlation coefficient of the linear fit, respectively.}
\label{cr}
\end{figure}

In Fig.\ref{compare}, the voids associated with star clusters or associations present lower intensity contrast and larger radius.
The anti-correlation between the intensity contrast and radius (Fig.\ref{cr}) suggests that voids with larger radii tend to have emptier central regions. 
This may be the result of more substantial feedback, or it could be due to the central regions of small-scale voids not being well resolved in the observations.

\subsection{Voids and their possible precursors within the network}\label{graph}

\begin{figure}
\centering
\includegraphics[width=0.5\textwidth]{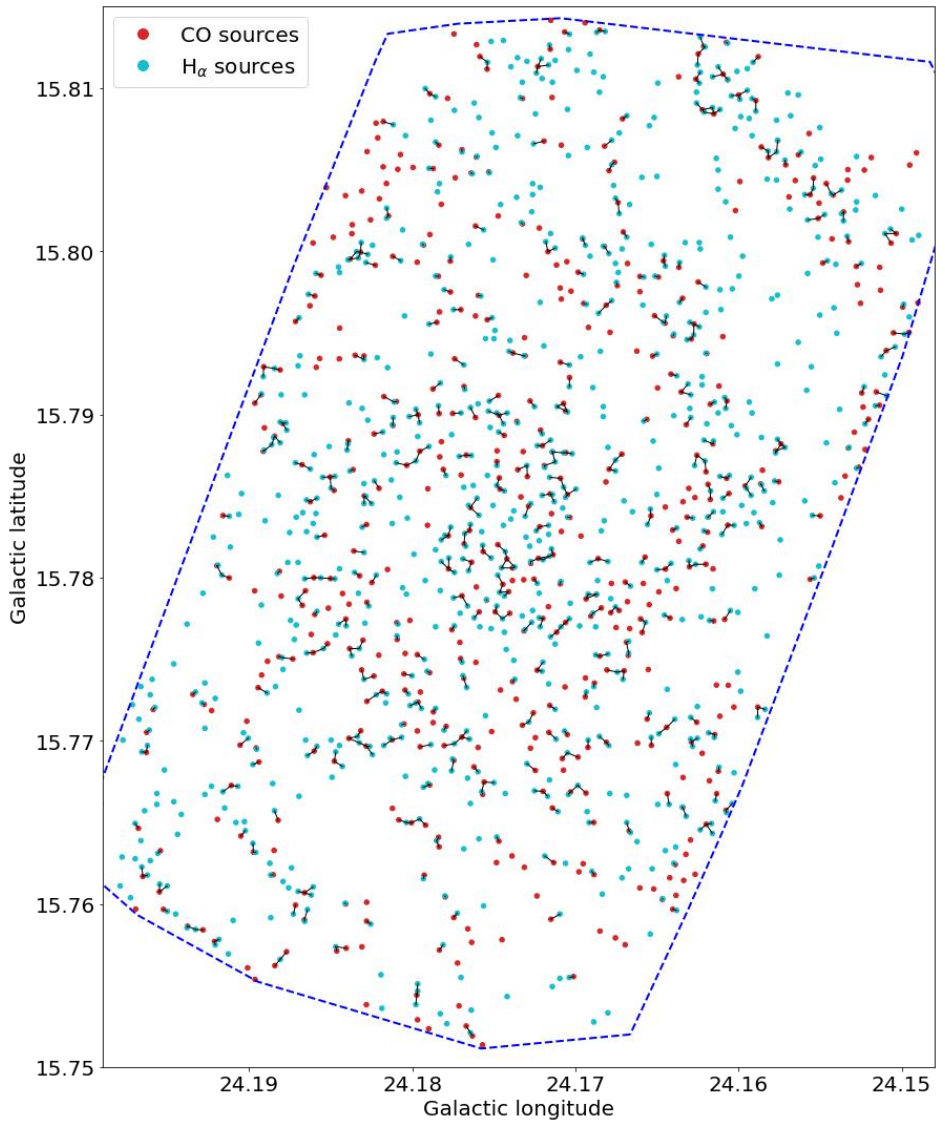}
\caption{The CO-H$_{\alpha}$ network constructed in Sec.\ref{graph}. Red and cyan dots represent CO and H$_{\alpha}$ sources, respectively. Blue dashed lines show the convex hull boundary determined by the spatial distribution of the 21$\mu$m sources.}
\label{network}
\end{figure}

\begin{figure*}
\centering
\includegraphics[width=0.7\textwidth]{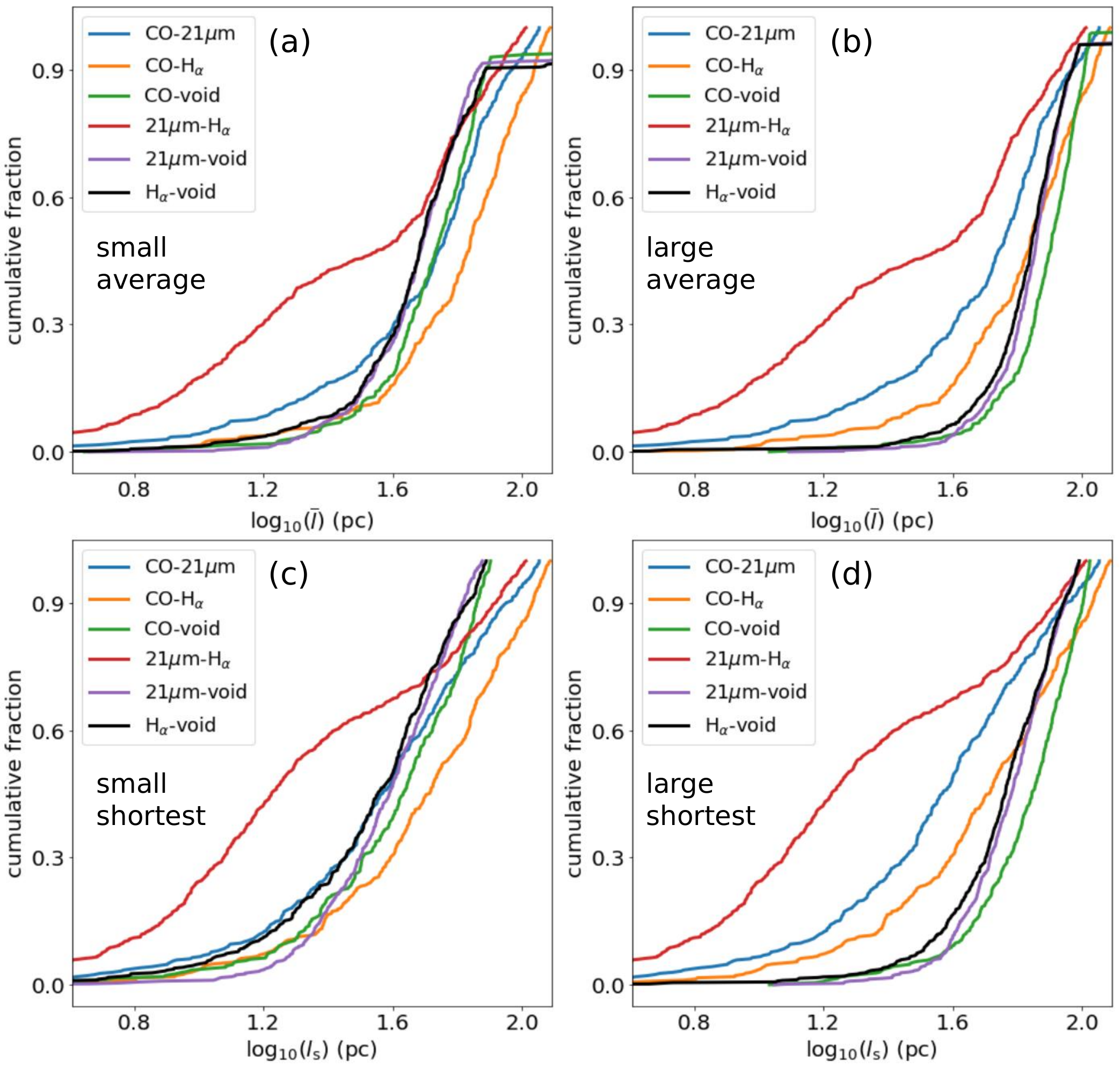}
\caption{
Distributions of average and minimum separations across all nine networks.
Using the CO–H$_{\alpha}$ network as an example, we examine all H$_{\alpha}$ sources neighboring a given CO source and calculate the average separation between them ($\overline{l}$), or focus on the nearest H$_{\alpha}$ source to determine the smallest separation ($l_{\rm s}$). The void sample is divided into two subsets based on the median void radius ("small" and "large").}
\label{sep}
\end{figure*}

In this section, we focus on the spatial relationships between voids and their possible precursors, as well as how these spatial relationships reflect their evolutionary (temporal) connections.

The studies by \citet{Kruijssen2014-439} and \citet{Kruijssen2018-479} introduced a statistically rigorous framework for translating the observed spatial offsets between cold gas and star formation rate (SFR) tracers into their corresponding evolutionary timescales. This methodology was applied by \citet{Kim2023-944} in an analysis of NGC 628, utilizing emission maps of CO, 21$\mu$m, and H$_{\alpha}$ to trace molecular clouds, embedded star formation, and exposed star formation. 
Since the progenitors of feedback-driven voids may be H$_{\alpha}$ sources (HII regions), could this methodology be further extended to include voids? The transition from CO to 21$\mu$m to H$_{\alpha}$ sources occurs within the same structure. However, once a H$_{\alpha}$ source evolves into a void, that original structure no longer exists. Therefore, the connection between a H$_{\alpha}$ source and a void involves two adjacent but distinct structures. The question now is whether the void is spatially closest to the H$_{\alpha}$ source compared to the CO and 21$\mu$m sources, given their potentially closer temporal association.

In \citet{Zhou2025-537-2630}, we modeled molecular gas in NGC 628 as a gravitational network, with molecular clouds represented as nodes. 
To systematically measure the separations between CO, 21$\mu$m, H$_{\alpha}$ sources, and voids, we construct a series of networks that link each pair of source types, called CO-21$\mu$m, CO-H$_{\alpha}$, CO-void, 21$\mu$m-H$_{\alpha}$, 21$\mu$m-void, and H$_{\alpha}$-void networks.
The elements of a network include nodes, edge weights, and connection rules. 
In this work, we focus solely on the spatial separations between nodes and therefore disregard the edge weights.
The nodes are the voids identified in this work and the CO, 21$\mu$m, and H$_{\alpha}$ sources identified in \citet{Zhou2024-534}. 
Compared to the ALMA and MUSE observations, the JWST observations have the smallest field of view. Therefore, we only consider the sources that fall within this smallest field of view. Specifically, we determine the convex hull boundary based on the spatial distribution of the 21$\mu$m sources (blue dashed lines in Fig.\ref{network}), and then retain only the sources that lie within this boundary.
Since we aim to examine the spatial separations between different types of sources, we need to exclude sources that are spatially isolated. This can be achieved by applying the connection rules.
We first utilized the minimum spanning tree (MST) method to link all the nodes in a network.
Then we impose a connection rule whereby nodes separated by a distance greater than the median edge length of the MST are not connected. We tested various percentiles of the MST edge length distribution and found that using the median strikes a balance—avoiding both excessive and insufficient connections—thus facilitating the examination of the neighboring sources. Moreover, connections are allowed only between different types of sources.

Fig.\ref{network} shows the CO-H$_{\alpha}$ network. This network structure defines the neighborhood relationships between CO and H$_{\alpha}$ sources. We can examine all H$_{\alpha}$ sources neighboring a given CO source and calculate the average separation between them ($\overline{l}$), or focus on the closest H$_{\alpha}$ source and determine the smallest separation ($l_{\rm s}$).
Assuming that voids undergo an evolutionary process from small to large sizes, we divide the void sample into two subsets based on the median void radius ($\sim$26 pc, "small" and "large"). By comparing these two subsets, we search for evidence supporting this hypothesis. The construction of the network is also carried out separately for the two subsets.
For all nine networks, their $\overline{l}$ and $l_{\rm s}$ are compared in Fig.\ref{sep}. 

21$\mu$m and H$_{\alpha}$ sources are closely associated in space. 
The notably shorter separation between them, relative to other pairs, suggests either a short embedded phase of very young stellar populations \citep{Kim2021-504, Kim2023-944}, or that the 21$\mu$m emission primarily arises due to heating from young stars in the HII region.
Compared to the small void group, for the large void group,
the separations between CO and 21$\mu$m, H$_{\alpha}$ and voids progressively increase, which is consistent with the evolutionary sequence in both time and space. Smaller voids are closer to molecular clouds, and as they grow larger, they move away from the clouds. 
A molecular cloud is considered associated with a void if the distance between their centers is smaller than the cloud radius.
Compared with molecular clouds not associated with voids, those associated with voids are significantly more massive.
As discussed in \citep{Zhou2024-534,Zhou2025-537-2630}, these more massive molecular clouds are also more evolved. 
In fact, 68\% of molecular clouds associated with voids are also associated with 21$\mu$m sources.
These results support a scenario in which some voids originate from molecular clouds.



\section{Conclusion}

A deep learning method (BlendMask) was used to identify voids in NGC 628 based on the PHANGS$-$JWST MIRI F770W image. To fully identify small-scale voids, we divided the image of the entire galaxy into 170 patches. We manually annotated voids in 10 independent patches located at different positions within the galaxy using LabelMe, and used them as the training set. BlendMask typically outputs segmentation masks for each detected object. We then fit ellipses to the boundaries of these masks.  
After eliminating duplicate detections, we introduced a void parameter (i.e. intensity contrast) to further refine the sample. 
Repeating the same procedure, we also used the bubble catalog of \citet{Watkins2023-944} to construct a training set to identify structures. In the end, we merged the structure catalogs obtained using these different methods, resulting in a final set of 5441 voids.

Voids are identified solely based on the F770W images. These voids may correspond to either feedback-driven bubbles or dynamically driven structures.
To further distinguish between them, we cross-matched the voids with star clusters and compact associations from archived catalogs. We found that only a small fraction (up to 17.6\%) of the voids are associated with these stellar populations.
An analysis of the HST B-band peak flux distribution for voids either associated or not associated with star clusters or associations reveals substantial overlap between the two groups. 
These results suggest that a significant number of star clusters and associations related to voids may remain unidentified or misclassified in existing catalogs. 
Voids associated with star clusters or associations tend to exhibit lower intensity contrast and larger radii. The observed anti-correlation between void radius and intensity contrast suggests that larger voids tend to have emptier centers, possibly due to more substantial feedback or limited resolution that blurs the centers of smaller voids.
Therefore, voids may serve as a complementary tool for identifying stellar populations and constraining their physical parameters.

The spatial relationships between voids and their potential precursors may reflect underlying evolutionary (temporal) connections. To systematically quantify the separations among CO, 21$\mu$m, H$_{\alpha}$ sources, and voids, we construct a series of networks that link each pair of source types. Across all nine constructed networks, we found that 21$\mu$m and H$_{\alpha}$ sources are most closely associated in space. 
Compared to the small-void group, for the large-void group,
the separations between CO and 21$\mu$m, H$_{\alpha}$ and voids progressively increase, which is consistent with the evolutionary sequence in both time and space. Smaller voids are closer to molecular clouds, and as they grow larger, they move away from the clouds. 
Compared with molecular clouds not associated with voids, those associated with voids are significantly more massive and appear to be more evolved. Actually, 68\% of molecular clouds associated with voids are also associated with 21$\mu$m sources.
These results support a scenario in which some voids originate from molecular clouds.

\section*{Acknowledgements}
We thank the referee for helpful comments that improved this work.
It is a pleasure to thank the PHANGS team. The data cubes and other data products provided by the team have greatly facilitated this work.
ALMA is a partnership of ESO (representing its member states), NSF (USA) and NINS (Japan), together with NRC (Canada), NSTC and ASIAA (Taiwan), and KASI (Republic of Korea), in cooperation with the Republic of Chile.

\section{Data availability}
All the data used in this work are available on the PHANGS team website
\footnote{\url{https://sites.google.com/view/phangs/home}}. 
The void catalog produced in this work is available via the MNRAS online supplementary materials.

\bibliography{ref}
\bibliographystyle{aasjournal}

\begin{appendix}

\section{Supplementary maps}

\begin{figure}
\centering
\includegraphics[width=0.5\textwidth]{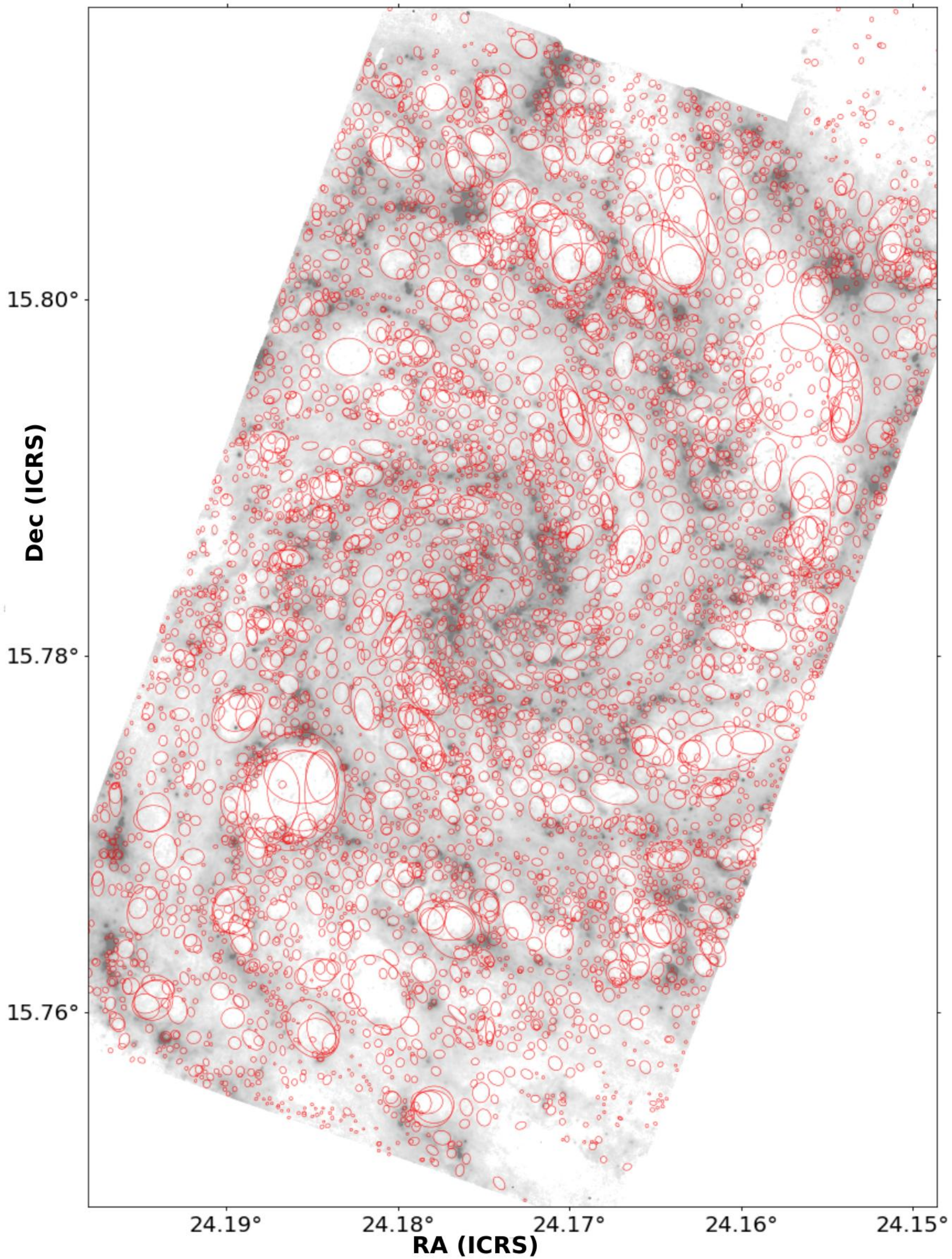}
\caption{Voids in the merged catalog after removing duplicate voids, as described in Sec.\ref{c}.}
\label{ML-merge}
\end{figure}

\begin{figure}
\centering
\includegraphics[width=0.5\textwidth]{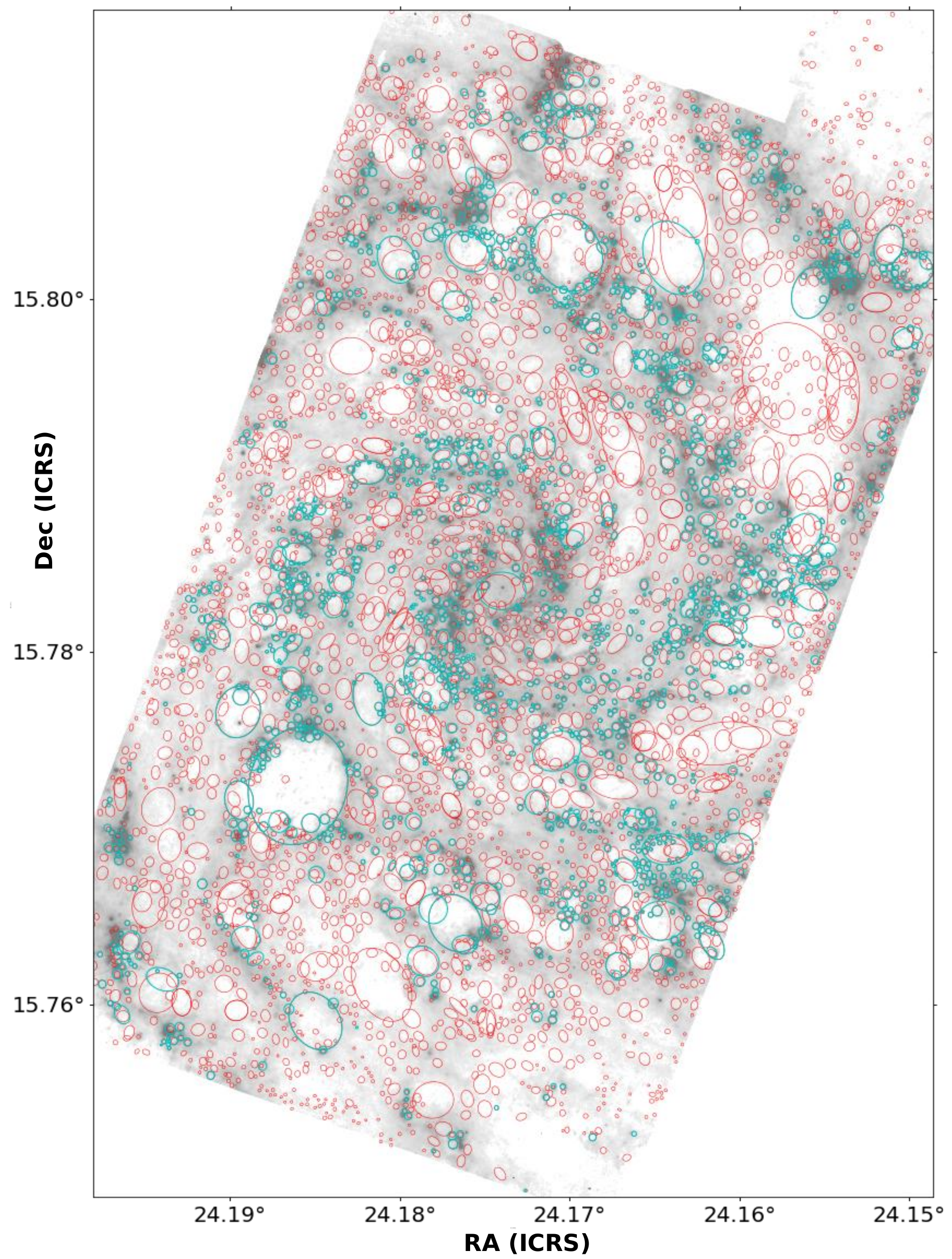}
\caption{A comparison between the 4594 voids retained after the final selection in Sec.\ref{layer} (also shown in Fig.\ref{ML-small}) and the bubble catalog of \citet{Watkins2023-944} (cyan). 94\% bubbles in \citet{Watkins2023-944} are included in the sample of 4594 voids, as discussed in Sec.\ref{c}.}
\label{ML-cp}
\end{figure}





\end{appendix}

\clearpage
\noindent
\end{document}